# Stokes Vector Scattering induced by Nonlinear Depolarization of Light in Fiber

*Lothar Moeller*

*SubCom, Eatontown, NJ 07716, USA, lmoeller@subcom.com*

**Keywords**: SOP, Stokes vector Scattering, Nonlinear depolarization, Comparative polarimetry.

## Abstract

We report the experimental observation of SOP scattering in Stokes space caused by nonlinear depolarization (NLDP) of light in fiber. In addition to our previous characterizations of NLDP by SOP speed histograms and Stokes vector spectra, we measure SOP scattering angles. For trans-Pacific propagation distances these angles range within a few degrees.

## 1 Introduction

Recently, a novel transmission phenomenon has been described; referred to as nonlinear depolarization of light (NLDP) in fiber[1]. Although relatively small in magnitude, its research work can contribute towards a more accurate understanding of transmission penalties for higher-order modulation formats.

As discussed in a previous work[1], un-polarized optical noise (ASE) rapidly changes the state of polarization (SOP) of a fully-polarized and co-propagating cw light by inducing antisymmetric phase noise in both of its orthogonal polarization states via the fiber's Kerr nonlinearity. These fluctuations become resolvable with a new generation of high-speed polarimeters and do not average out over wide noise bandwidths, but instead grow with propagation distance. Our recent reports include the experimental time domain observation of NLDP by means of SOP speed histograms[2] and in the frequency domain by using Stokes vector spectroscopy[3]. A third approach to illustrate NLDP is based on specifying an uncertain region on the Poincare sphere for scattered SOPs. During long propagation, NLDP rapidly deflects the SOP of a cw probe from a center position within a small solid angle. While at any instance of time, a single point on the Poincare sphere represents the probe's SOP, we can imagine the NLDP-induced SOP deflections as SOP blurring, i.e. within a small area around its center position, the SOP performs a kind of fast random walk with substantial path changes on a ns-scale. For fast SOP changes beyond the temporal resolution capabilities of deployed measurement equipment, such fluctuations are more suitable characterized as a steady SOP uncertainty or blurring. The quantities of SOP speed histograms, Stokes vector spectra, and the SOP blurring are interrelated. Knowledge of two allows estimates of the third.

While other kinds of SOP blurring have been previously discussed in the context of fiber communications[4], we emphasise that NLDP is unrelated to those and should especially not be confused with the similar-sounding term of nonlinear polarization rotation of light[5]. NLDP depends on PMD, whereas aforementioned effects do not.

Here we experimentally determine an average solid angle on the Poincare sphere containing the scattered SOPs.

A major challenge in measuring the NLDP-induced SOP scattering results from noise and drift caused by environmental factors like temperature fluctuations and vibrations of the required ultra-long transmission path. However, our data processing selects and filters the fast NLDP-induced SOP scattering from other slower environmental factors.

## 2 Comparative Polarimetry sorts out Non-NLDP-related SOP Noises

NLDP is a relatively small NL phenomenon that builds up over several thousand km of transmission length, during which it will get masked by noise from EDFAs. Specifically, the degraded OSNR of a probe tone produces dominant signal-ASE beating during the detection process which hides NLDP. Other noises like ASE-ASE beat noise, thermal noise, or A/D quantization noise in the polarimetric detection generate additional artefacts to which we refer hereafter as NIASC (**N**oise **I**nduced **A**rtificial **S**OP **C**hanges). In general, any noise that impacts the recording of Stokes parameters from a probe leads to a tangential motion of the normalized Stokes vector on the Poincare sphere and obscures SOP speed and scattering. The concept of comparative polarimetry[1] compares the outputs of a transmission and a back-to-back (btb) measurement to visualize NL propagation effects. In the first experiment a probe undergoes both NLDP during long distance propagation and NIASC through its O-E conversion. In the subsequent btb launch, the probe is impacted by identical NIASC (ASE-signal beat noise, thermal noise, etc.) but not by NLDP. Optical 'fingerprints' of a signal like SOP speed histograms, Stokes vector spectra, or SOP scattering discussed herein differ for both experiments and reveal NLDP. In case of transmission, the probe's inherent SOP drift due to temperature fluctuations, motion and vibrations of the fiber could be interpreted as NLDP and thus requires data post-processing to elucidate the true signal.

## 3 Data Post-Processing Cancels SOP Drift for an Unbiased Comparative Polarimetry

The SOP drift impacts the polarimetric detection in a fundamentally different way than the aforementioned stationary noise processes (e.g. signal-ASE beat noise). Those can be eliminated by sufficiently long integration times. For a SNR-limited probe after transmission but without SOP drift, a reduction of the detector bandwidth will lead in principle, to a single-point representation of its SOP on the Poincare sphere. In contrast, biasing of data caused by the probe's SOP drifting across the Poincare sphere, cannot be mitigated with a reduced detection bandwidth. SOP drift during transmission is unavoidable; stemming mainly from temperature fluctuations and mechanical vibrations of the test bed fibers. Since the time scales underlying these three processes, (SOP drift, NLDP, and NIASC), differ significantly, they can be separated during the data post-processing.

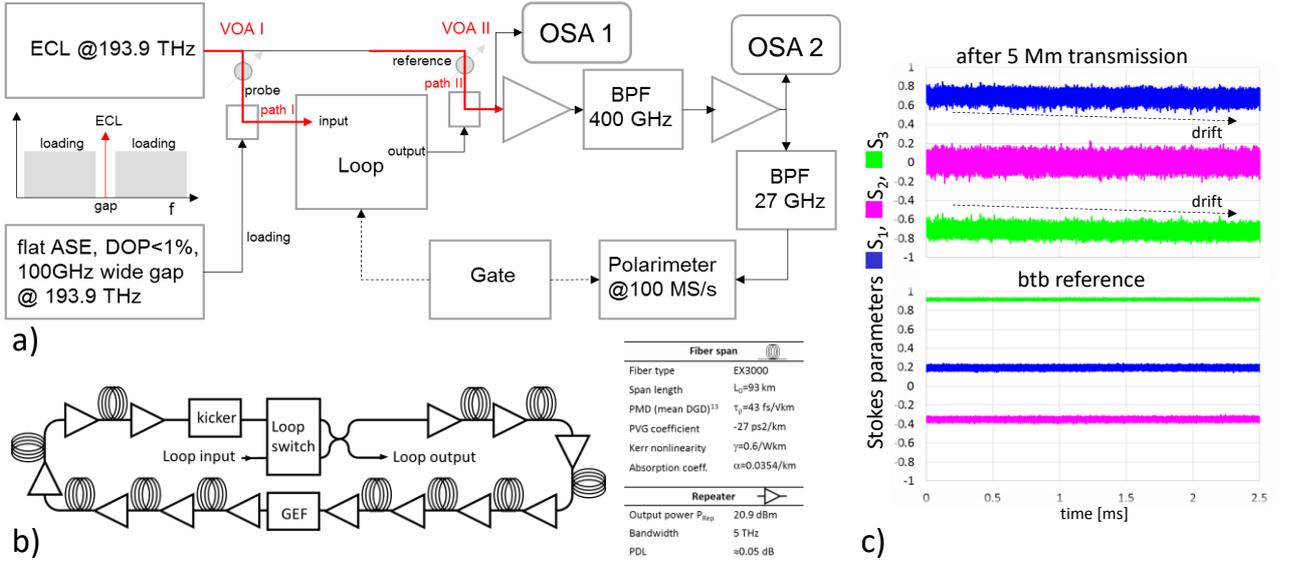

**Fig.1: Measurement of NLDP-induced SOP scattering after ultra-long haul transmission.**
a) A high-speed polarimeter records Stokes vectors of a probe propagating either through a recirculating fiber loop (RFL, path I) or btb (path II). Both results enable comparative polarimetry. b) The RFL comprised of 11 identical spans allows for adjustable signal propagation lengths and repeater output powers to study the dependence of NLDP on the transmission distance and the Kerr nonlinearity.
c) Stokes parameters of probe within one sampling window after 5 Mm transmission and in btb. Widen traces of transmitted probe indicate SOP scattering. Small low speed drift of its $S_1$ and $S_3$ is visible.

The data processing outlined below, applies sufficiently long low-pass integration to determine center positions of a hypothetically-assumed undistorted SOP; but short enough so SOP drift can be neglected within the integration time. We record the four instantaneous components of a normalized Stokes vector over a period $\tau_{rc}$ of 2.6 ms and split the string into $N$ shorter intervals $\tau_{avg\_i}$ of about 10 us length. For these we compute average *normalized* Stokes vectors that described the position of an undistorted SOP. This parameter choice in the analysis eliminates vibrational impact of up to ~10 kHz but also allows to capture the majority of the probe's Stokes vector spectrum which extends into the several MHz range[3]. We define the spot size of the presumed small SOP scattering by relating the Stokes vector variance (Eq.1) to a corresponding solid angle in Stokes space. The difference between the two solid angles $\Omega_{trans}$ and $\Omega_{btb}$ taken for the comparative polarimetry yields the NLDP-induced SOP scattering $\Omega_{NLDP}$ (Eq.2)

$$\Omega_{trans}_{(btb)} = \frac{\pi}{N}\sum_{i=1}^{N}\frac{1}{\tau_{avg\_i}}\int_{\tau_{avg\_i}}[\overrightarrow{S(t)}-\frac{1}{\tau_{avg\_i}}\int_{\tau_{avg_i}}\overrightarrow{S(t)}dt]\big|_{norm}^2 dt \quad (1)$$

$$\Omega_{NLDP} = \Omega_{trans} - \Omega_{btb}, \quad (2)$$

where $\overrightarrow{S(t)} = (S_1, S_2, S_3)$ and $|_{norm}$ stand for the 3 dimensional Stokes vector of the probe and the normalization of the averaged vectors, respectively. Decreasing the polarimetric detection bandwidth by increasing its integration time $\tau_{dec}$ (in typical experimental conditions) leads to a reduction in NIASC, proportional to $\tau_{dec}^{-2}$. But this also limits the detector's ability to track fast NLDP-induced SOP changes and enhances the measurement error for $\Omega_{NLDP}$ to $\sim \pi/2\ (v_{D\_SOP}\ \tau_{dec})^2$ where $v_{D\_SOP}$ is the speed of a slow SOP drift. Our measurement method avoids such trade-offs and allows for fast SOP tracking ~100Mrad/s due to ~30 MHz detector bandwidth, NIASC elimination by comparative polarimetry, and SOP drift offsetting based on short term averaging during the NL data post-processing.

## 4 Test Bed Operation for Assessing SOP Scattering in long-haul transmission

For typical system parameters, NLDP appears at an experimentally accessible magnitude after several thousand km propagation. Recirculating Fiber Loops (RFL) emulate long-haul propagation over various distances and at different span launch powers to study the NL dependences of SOP scattering (Fig.1). In contrast to real systems, the signal's periodic transit through the same fiber spans leads to its polarization-dependent filtering (PDL, PMD). To randomize such effects and approach a more realistic propagation characteristic, an active loop synchronous polarization controller[6] (Kicker) transforms the SOP of an incoming signal to a new and randomly chosen output SOP. After each loop roundtrip of the signal the kicker is reprogrammed but stays stationary for the duration of the passing signal so no SOP drift is introduced.

A network embedding the RFL enables switching between transmission and btb operations required for comparative polarimetry. The probe enters the RFL via VOA_1 while VOA_2 blocks a reference path (Fig.1). On the receive side, a gated polarimeter detects the optically amplified and narrow bandwidth filtered probe. For the btb measurement VOA_1 prevents the probe from entering the RFL while it passes VOA_2 and combines with repeater noise stemming from the RFL. In both measurements transmission and btb, the VOAs adjust identical OSNRs and signal powers on the OSA_2 (Fig.1). The setup is similar to test beds previously used for recording NLDP-induced SOP speed[2] and Stokes vector spectra[3], but differs in the polarimeter operation. A gated polarimeter, synchronized with the RFL output, samples at 100 MS/s and 14 bit resolution the probe's Stokes vector over a window of 2.6 ms duration. The raw data are written real-time into RAM and off-line post-processed. As the NLDP-induced SOP scattering depends on the individual arrangement of the fiber birefringence along the transmission path, we analyse several dozen gate windows and report their averages. For each gate window the probe's launch SOP into the RFL and the sequence of the SOP transforms performed by the kicker are different to emulate various assembles of birefringence.

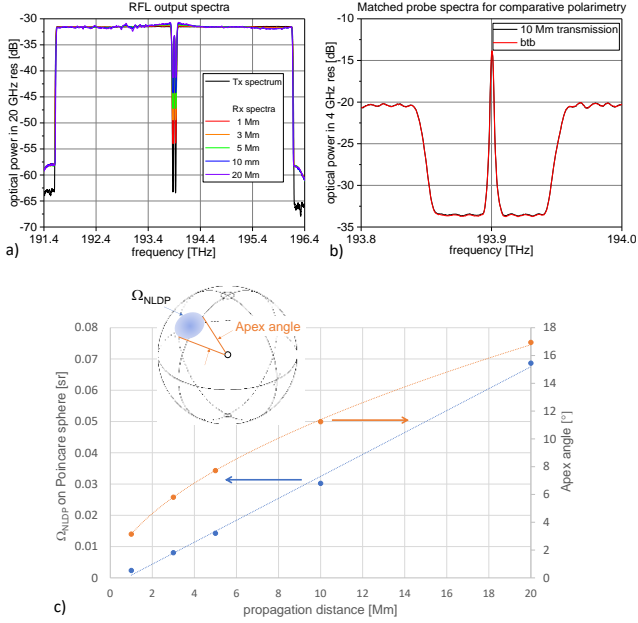

Fig.2: a) The Tx spectrum and RFL output spectra at OSA 1 after 1 to 20 Mm propagation maintain their shapes despite noise level shifts within their centre gap. b) OSA 2 recorded probe spectra after 10 Mm transmission and of the corresponding btb measurement show almost perfect matching. c) NLDP-induced SOP scattering vs propagation distance described by solid angles on the Poincare sphere and the corresponding apex angles.

## 5 SOP Scattering grows with Propagation distance and Repeater power of Setup

To simplify a theoretical analysis, we chose less complex experimental conditions and adjust a flat gain shape in our test bed. At launch, there is a cw probe at 193.9 THz in a 100 GHz-wide gap within a boxcar-shaped ASE spectrum. These experimental conditions are used for propagation distances of 1 to 20 Mm and produce similar receive spectral shapes (Fig.2a). Spectral matching of the transmitted probe and its comparative btb signal could be achieved with better than 0.1 dB accuracy as exemplified in Fig.2b. Solid angles for NLDP-induced SOP Scattering (defined by Eq.2) were recorded at 1, 3, 5, 10, and 20 Mm propagation lengths and grow monotonously over distance (Fig.2c). We additionally list the corresponding apex angles of the solid angles which are probably easier to envision.

To rule out vibrations or temperature fluctuations of the transmission path as dominant cause for the distance dependent SOP spot sizes, we decrease in two measurements the repeater output power in every loop span by 1 dB and 2 dB, respectively. While reduced launch power into a span weakens the Kerr nonlinearity and therefore lessens NLDP-induced SOP scattering, other environmental factors that can generate SOP fluctuations are unchanged. At 10 Mm transmission we measured for the 3 repeater output powers (20.9 dBm, 19.9 dBm, and 18.9 dBm) significantly decreasing SOP scattering (table in Fig.3). NLDP decreases with increasing frequency spacing between the probe and a spectral component of the ASE loading[1]. However, as the spectral envelope of the loading bends convexly at reduced repeater power (Fig.3), these measurements show that the detected SOP scattering results from a NL process and not from vibrations or temperature changes of our setup. Note, if the ASE spectrum would have been instead concavely shaped, the previous conclusion could not be drawn without detailed analysis.

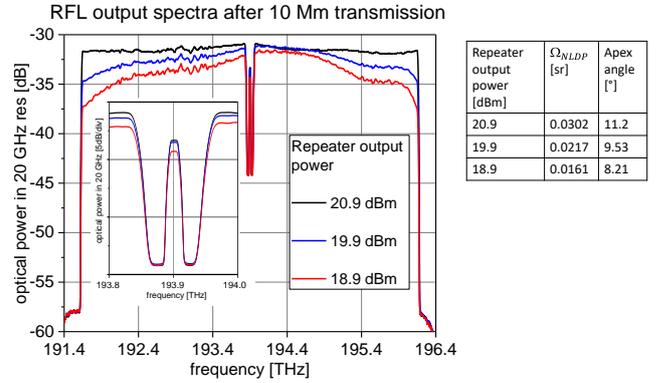

| Repeater output power [dBm] | $\Omega_{NLDP}$ [sr] | Apex angle [°] |
|---|---|---|
| 20.9 | 0.0302 | 11.2 |
| 19.9 | 0.0217 | 9.53 |
| 18.9 | 0.0161 | 8.21 |

Fig.3: Receive spectra after 10 Mm propagation for three different repeater output powers. Convex bending of the spectra at lower repeater power is visible. SOP scattering decreases at smaller repeater output powers.

## 6 Conclusions

We have successfully demonstrated NLDP-induced SOP scattering. Owing to NL interactions with an unpolarized and co-propagating ASE, the received SOP of a cw probe fluctuates within a small spot size on the Poincare sphere. A corresponding distance-dependent apex angle amounts to a few degrees over trans-Pacific distances. Other environmental factors such as vibrations were ruled out as source of the observed SOP fluctuations by varying the magnitude of the fiber nonlinearities. Comparative polarimetry combined with data post-processing unambiguously reveals the SOP scattering by NLDP.

## 6 References

[1] L. Moeller, APL Photonics Vol.5, 10.1063, 2020 (arXiv:1902.05168, 2019).
[2] L. Moeller, ECOC'19, Tu.1.C.4, 2019.
[3] L. Moeller, ECOC'19, Tu.1.C.5, 2019.
[4] R. Dar et al., ECOC'16, p. 483-6, 2016.
[5] B.C. Collings et al., IEEE Photon. Technol. Lett. 12, pp. 1582-4, 2002.
[6] Q. Yu et al., J. Lightwave Technol. 21, pp. 1593-1600, 2003.